\documentclass[a4paper,11pt,showkeys]{revtex4}
\usepackage[utf8]{inputenc}
\usepackage[T1]{fontenc}
\usepackage{lmodern}
\usepackage{epsfig}
\usepackage{array}
\usepackage{amsmath}
\usepackage{longtable}
\usepackage{epstopdf}
\usepackage{graphicx}
\graphicspath{/}
\usepackage{tabularx}

\begin{document}

\title{Study of damped oscillating structures from charged and neutral K-meson electromagnetic form factors data}

\date{\today}

\medskip

\author{Stanislav Dubni\v cka}
\address{Institute of Physics, Slovak Academy of Sciences, Dubravska cesta 9,
SK-84511 Bratislava, Slovak Republic}

\author{Anna Zuzana Dubni\v ckova}
\address{Department of Theoretical Physics, Comenius University,Mlynska dolina,
SK-84248 Bratislava, Slovak Republic}

\author{Luk\' a\v s Holka}
\address{Institute of Physics, Slovak Academy of Sciences, Dubravska cesta 9,
SK-84511 Bratislava, Slovak Republic}

\author{Andrej Liptaj}
\address{Institute of Physics, Slovak Academy of Sciences, Dubravska cesta 9,
SK-84511 Bratislava, Slovak Republic}

\begin{abstract}
     The damped oscillating structures were recently revealed in the proton "effective" form factor data. For the time being they
can be neither confirmed nor disproved by investigations of timelike data on the individual proton electric and proton magnetic form factors
because their precision and reliability (especially of the proton electric form factor data) has not achieved required level for this aim.

     On the other hand, conjectures that the damped oscillating structures are direct manifestations of the quark-gluon structure of the proton indicate that
they must not be specific only for the proton and neutron, but that they should be present also for other hadrons. This opens a plausibility to find damped oscillatory structures also from the electromagnetic form factors data of such hadrons, for which adequate electromagnetic
form factors data exist, by using the same procedure as for the proton. Consequently in this paper damped oscillatory structures are investigated in the electromagnetic form factors data of the charged and neutral $K$-mesons to be extracted from the corresponding production cross sections, $\sigma^{bare}_{tot}(e^+e^-\to K^+ K^-)$  measured from the threshold up to 64 GeV$^2$ and $\sigma^{bare}_{tot}(e^+e^-\to K_s K_L)$ measured from the threshold up to 9.5 GeV$^2$ of the total c.m.
energy squared.

     The following results have been obtained. If the charged and neutral K-meson electromagnetic form factors timelike data are described by the three parametric
formula by means of which damped oscillating structures have been revealed from the "effective" proton form factor data, with, however, rather large value of $\chi^2/ndf$. Then oscillating structures appear. If physically well founded Unitary and Analytic model of the K-meson electromagnetic structure is used for a description of the charged K-meson electromagnetic form factors timelike data, no damped oscillating structures are visible. However, in the case of the neutral K-meson electromagnetic form factor timelike data one cannot make a definite decision, especially in the region of the $\phi$-resonance peak where
new more precise data are indispensable.

     The results on the charged and neutral K-meson electromagnetic form factors data indicate that damped oscillating structures obtained
from the "effective" proton form factor data are likely artefact of the three parametric formula which does not describe these
data with a sufficient precision.

\end{abstract}


\maketitle
\section{INTRODUCTION}
   There have been historically two different approaches used in measurements of the total cross section $\sigma^{bare}_{tot}(e^+e^- \to p \bar p)$.
The scan method \cite{Pedlar, Ablikim1, Ablikim2, Akhmetshin1, Akhmetshin2, Ablikim3} in which the measurement of $\sigma^{bare}_{tot}(e^+e^- \to p \bar p)$ is carried out by a change of the electron-positron collider energy from one value to another, and the novel more precise measurements by the initial state radiation (ISR) technique of the process $e^+e^-\to p\bar p(\gamma)$ \cite{Aubert, Lees1, Lees2, Ablikim4, Ablikim5} in which the c.m. energy squared $s$ of the collider is fixed at the value of the maximal collider luminosity with subsequent measurement of $\sigma^{bare}_{tot}(e^+e^- \to p \bar p)$ by a measurement of energy of the radiated photon from initial electron or positron. In this way values of $\sigma^{bare}_{tot}(e^+e^- \to p \bar p)$ are obtained from the threshold up to the energy at which the collider achieves the maximal value of its luminosity. This technique gives the data on $\sigma^{bare}_{tot}(e^+e^- \to p \bar p)$ with reduced errors in comparison with the scan method data
and consequently in our investigations the data obtained by the ISR technique are always preferred.

   The behavior of measured $\sigma^{bare}_{tot}(e^+e^- \to p \bar p)$ is described theoretically by the relation

\begin{eqnarray}\label{totcspp}
 \sigma_{tot}(e^+e^- \to p \bar p)=\frac{4 \pi \alpha^2 C_p \beta_p(s)}{3 s}
 [|G^p_M(s)|^2+\frac{2m_p^2}{s}|G^p_E(s)|^2],
\end{eqnarray}
with $\beta_p(s)=\sqrt{1-\frac{4 m^2_p}{s}}$, $\alpha$=1/137 and where $C_p=\frac{\pi \alpha / \beta_p(s)}{1-\exp(-\pi \alpha / \beta_p(s))}$
is the so-called Sommerfeld-Gamov-Sakharov Coulomb enhancement factor \cite{BaPaZa}, which accounts for the electromagnetic (EM) interaction between the outgoing proton and antiproton, and $s$ is the c.m. energy squared.

   The functions $G^p_E(s)$ and $G^p_M(s)$ in Eq. (\ref{totcspp}) are the Sachs proton electric and proton magnetic form factors (FFs), respectively.
At the proton-antiproton threshold the two FFs have the same value as follows from their definition through the Dirac and Pauli proton FFs.

   As one cannot determine both proton EM FFs, $G^p_E(s)$ and $G^p_M(s)$, from measured $\sigma^{bare}_{tot}(e^+e^- \to p \bar p)$ at a given value
$s>4m^2_p$ simultaneously, experimental groups \cite{Lees1, Ablikim2, Ablikim4, Ablikim3, Ablikim5}, with the hope of getting more information about the proton structure, generalized the threshold identity  $|G^p_E(4m^2_p)|\equiv|G^p_M(4m^2_p)|$ in Eq. (\ref{totcspp}) for all higher $s$-values up to $+\infty$, and the data with errors were obtained by means of the following expression
\begin{eqnarray}
  |G^p_{eff}(s)|=\sqrt{\frac{\sigma^{bare}_{tot}(e^+e^- \to p \bar p)}{\frac{4\pi\alpha^2C_p\beta_p(s)}{3s}(1+\frac{2m^2_p}{s})}}
\end{eqnarray}
which defines the the so-called "effective" FF.

   Immediately after the BABAR data on $G_{eff}$ were published \cite{Lees1, Lees2}, the modified form \cite{TomRek} of the dipole
formula for nucleon EM FFs behavior in the spacelike region
\begin{eqnarray}\label{fortfunct}
  G^p_{eff}=\frac{A}{(1+s/m_a^2)(1-s/0.71 GeV^2)^2},
\end{eqnarray}
with parameter values $A=7.7$ and $m^2_a=14.8$ GeV$^2$ (without quoting neither the parameter errors nor the value of $\chi^2/ndf$) has been applied successfully for their description. Afterwards, the result of the best fit given by  (\ref{fortfunct}) has been subtracted from BABAR data \cite{Lees1, Lees2} on the "effective" FF with errors and in the plot of these differences damped oscillatory structures with regularly spaced maxima and minima have been revealed \cite{BianTom}. They were shown as  function of $p(s)=\sqrt{s(\frac{s}{4m^2_p}-1)}$ which corresponds to the three momentum of one of the proton or antiproton in the frame where other one is at rest.

   We have collected all existing data on $G^p_{eff}$ \cite{Lees1, Lees2, Ablikim2, Ablikim4, Ablikim3, Ablikim5} obtained till now, repeated their analysis
with slightly different parameters with errors $A=8.9\pm0.3$ and $m^2_a=9.2\pm0.8$ GeV$^2$ and confirmed minima and maxima from \cite{BianTom}, see Fig.1.
We show the structure as a function of $s$, the $\chi^2/ndf\approx 5$ is very high, meaning that from the statistic point of view data are not well described.
\begin{figure}[ht]
  \includegraphics[width=55mm]{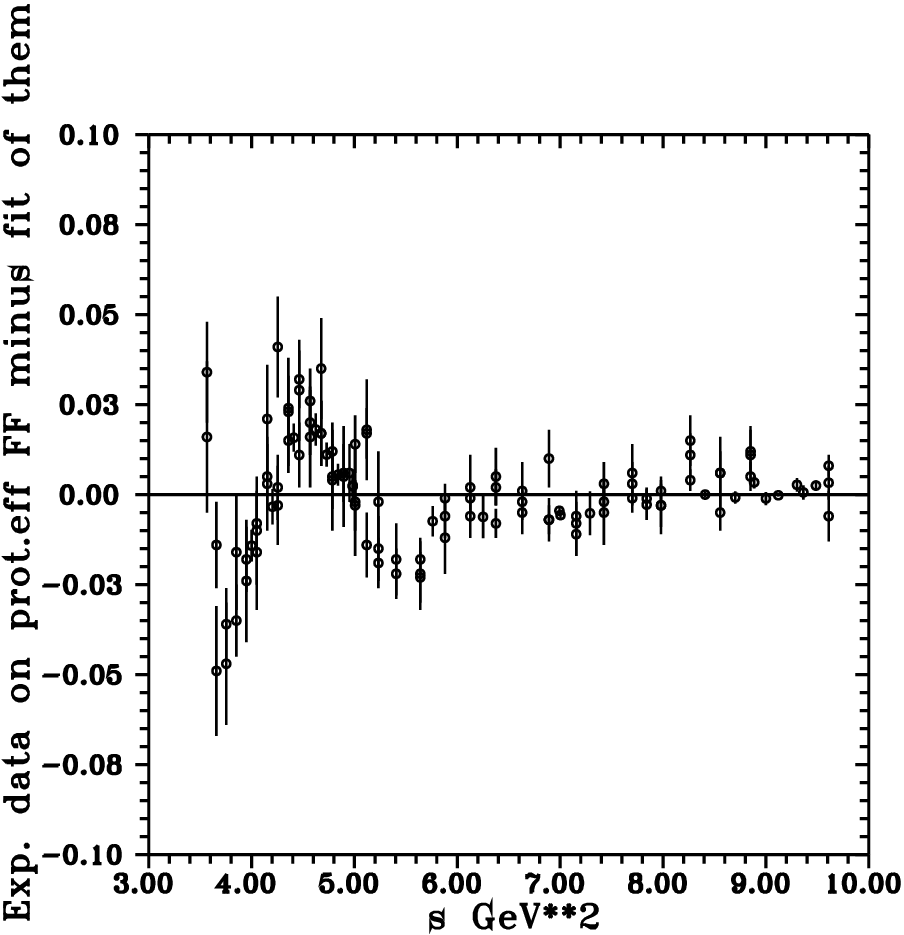}\\
\caption{Result of a subtraction of the best fit of all existing data on the proton "effective" form factor from these data with errors.}\label{FIG.1:}
\end{figure}

   The numerical value of $\chi^2/ndf$ is not quoted in the paper \cite{BianTom}, but we suppose that it is also not acceptable to claim a good description of
the investigated data. Therefore, we presume that the damped oscillatory structures are simply an artefact of an insufficiently precise fit by (\ref{fortfunct}) and does not reflect any property of the proton structure.

   If this is incorrect, the origin of the damped oscillatory structures from the proton "effective" FF data remains unknown.

   However, there are conjectures \cite{T-GP} that the damped oscillatory structures are direct manifestation of the quark-gluon structure of the proton.
If so, then they can not be specific only to the proton and neutron, but should be present also for other hadrons.

   Nowadays there are three candidates of other hadrons for which sufficient reliable EM FF data for such investigations exist and simultaneously there is also a
physically well founded theoretical model for their accurate description. These are the charged pion and the charged and neutral K-mesons.

   The search for the damped oscillating structures from charged pion EM FF data has been carried out in the recent paper \cite{BDD}. Here
with an aim to acquire more concrete support for the latter idea, damped oscillatory structures from the charged and neutral
K-meson  EM FFs timelike data are investigated.

\section{INVESTIGATION OF DAMPED OSCILLATING STRUCTURES FROM CHARGED K-MESON EM FF TIMELIKE DATA}

   The timelike data on the charged K-meson EM structure are contained in the measured total cross section
$\sigma^{bare}_{tot}(e^+e^- \to K^+ K^-)$. To evaluate these data from $\sigma^{bare}_{tot}(e^+e^- \to K^+ K^-)$, no extra nonphysical assumptions are needed, unlike in the case the nucleon "effective" FF data from $\sigma^{bare}_{tot}(e^+e^- \to p \bar p)$, because there is only one function $|F_K^{\pm}(s)|$ completely describing the measured total cross section of the electron-positron annihilation into $K^+K^-$ pair. The $|F_K^{\pm}(s)|$ with errors is calculated by means of the following relation

\begin{eqnarray}\label{chKFF}
  |F_{K^{\pm}}(s)|=\sqrt{\sigma^{bare}_{tot}(e^+e^- \to K^+ K^-)\frac{3s}{\pi\alpha^2C_K{^{\pm}}\beta^3_K{}^{\pm}}}
\end{eqnarray}
with $\beta_K{^\pm}(s)=\sqrt{1-\frac{4 m^2_K{^\pm}}{s}}$, $\alpha$=1/137 and where $C_K{^\pm}=\frac{\pi \alpha / \beta_K{^{\pm}}(s)}{1-\exp(-\pi \alpha / \beta_K{^{\pm}}(s))}$ is the so-called Sommerfeld-Gamov-Sakharov Coulomb enhancement factor \cite{BaPaZa} of charged kaons, which accounts for EM interaction between the outgoing $K^+K^-$. The total cross section data we use in (\ref{chKFF}) will be taken from two recent ISR measurements of the process $e^+e^- \to K^+ K^-(\gamma)$, one  from \cite{Lees3} for $s<25$ GeV$^2$, and another from  \cite{Lees4} in the range 6.76 GeV$^2$ $<s<$ 64 GeV$^2$  and also from the measurement \cite{Ablikim6} for 4.0 GeV$^2$ $\leq s \leq $ 9.49 GeV$^2$, which achieved best precision for the process $e^+e^- \to K^+ K^-$ at a number of c.m. energies by the scan method with the BESSIII detector at BEPCII. All these three data sets are graphically presented in Fig.2
\begin{figure}
    \includegraphics[width=0.30\textwidth]{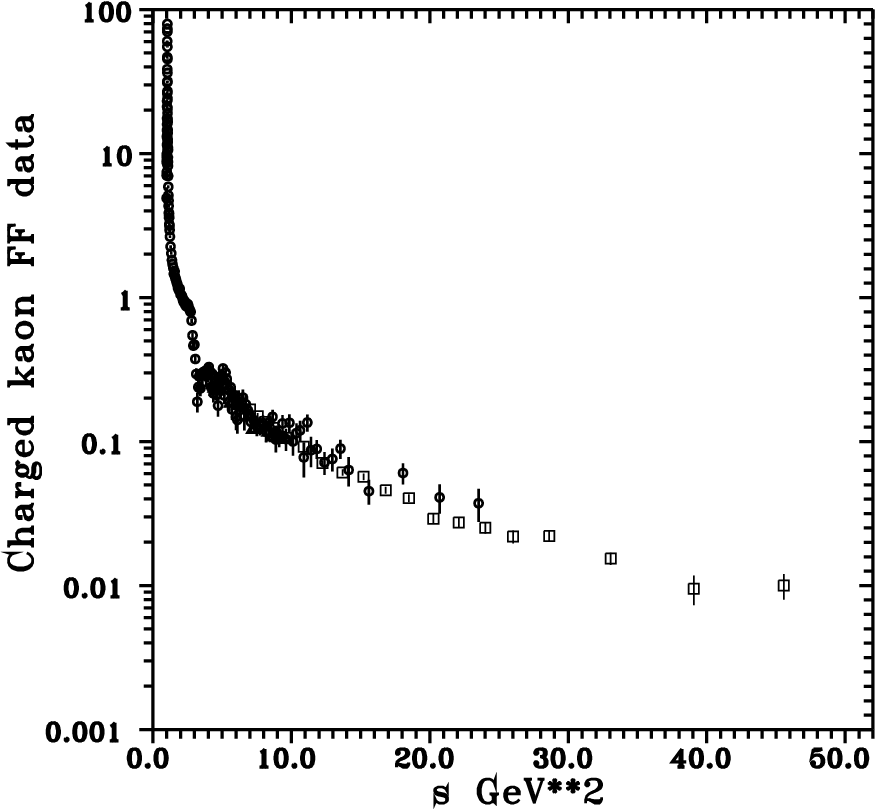}\hspace{0.3cm}
    \includegraphics[width=0.30\textwidth]{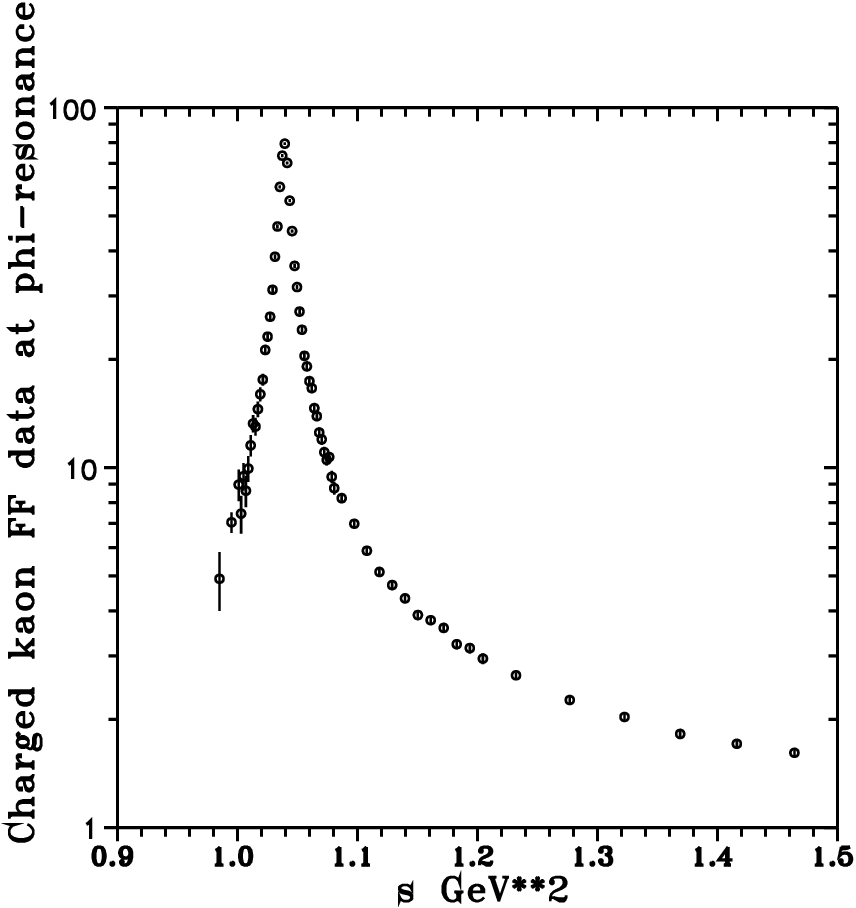}\\
\caption{Charged kaon EM FF data $|F_K^{\pm}(s)|$.}\label{chkdat}
\end{figure}
and the region (2-7) GeV$^2$ in more detail shown in Fig.3.

\begin{figure}[bth]
  \includegraphics[width=.30\textwidth]{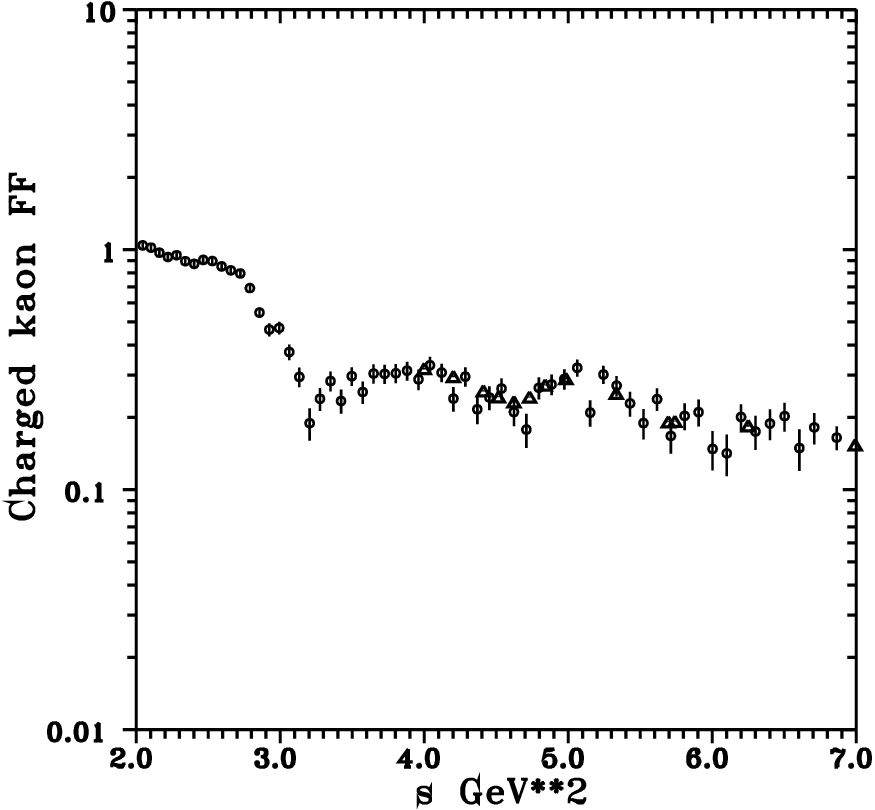}\\
  \caption{The data on $|F_{K^{\pm}}(s)|$ at the region 2-7 GeV$^2$ in more detail.}\label{FIG. 3:}
\end{figure}

   As one can see from Fig.2, the data in \cite{Lees3} above 6.5 GeV$^2$ are very scattered, even in some points inconsistent. As a result it is impossible to achieve
a statistically acceptable $\chi^2/ndf$ value in the description of such data. Fortunately, the same experimental BABAR group in the paper \cite{Lees4} repeated measurements of the $e^+e^- \to K^+ K^-$ process from 6.76 GeV$^2$ to 64 GeV$^2$ and obtained more precise and reliable data. Hence we have excluded all data from \cite{Lees3} in the energies 6.76 GeV$^2$ - 25 GeV$^2$ and substituted them by more precise data from \cite{Lees4}. Moreover, we also did not consider the scan data measured by BESIII collaboration \cite{Ablikim6} in the range (4-9.49) GeV$^2$ because they are in tension with the selected data from \cite{Lees3} and \cite{Lees4}.

   In order to investigate possible oscillatory structure from the charged K-meson EM FF timelike data by using the same procedure as for the proton, the modification
of the formula (\ref{fortfunct}) has been made, in the sense that the magic nucleon number 0.71 GeV$^2$ was left as a free parameter A3 in our analysis. Then the best description of the data in Fig.2 is achieved with A3=$0.8403\pm0.0024$ GeV$^2$, $m^2_a$=$0.2400\pm0.0709$ GeV$^2$ and A=$5.14773\pm0.0013$, as it is graphically presented in Fig.4 by the dashed line. If dashed curve data are subtracted from selected charged K-meson FF data with errors damped oscillating structures are observed around the line crossing the zero as seen in Fig.5.
\begin{figure}
    \includegraphics[width=0.30\textwidth]{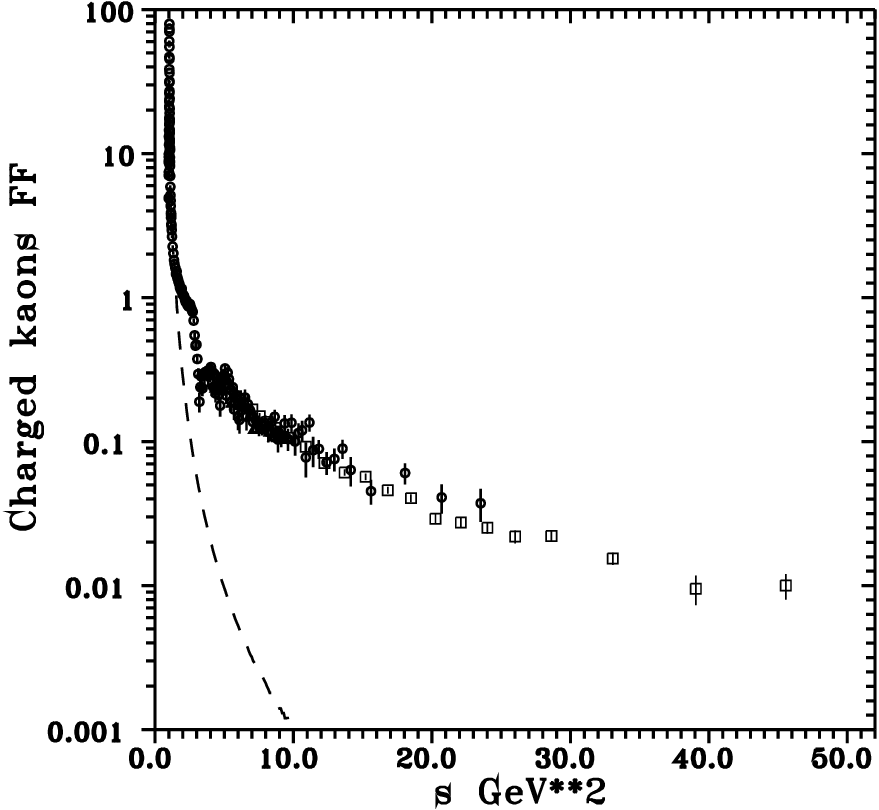}\hspace{0.3cm}
    \includegraphics[width=0.30\textwidth]{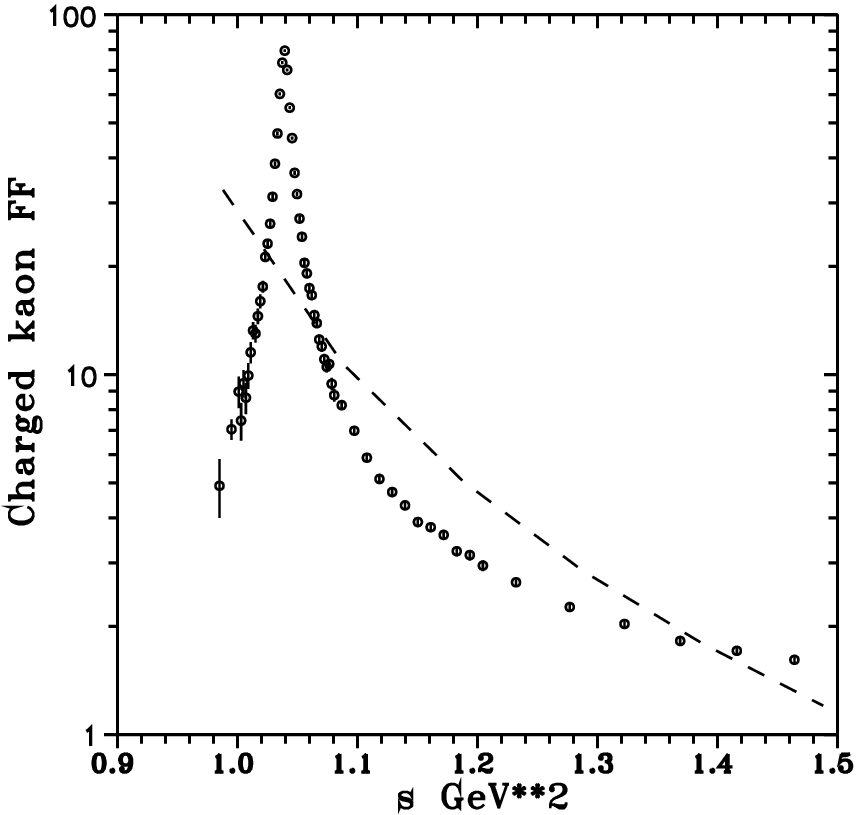}\\
\caption{Charged kaon EM FF data described by dashed line as generated by the modified three parametric formula (\ref{fortfunct}).}\label{chkdat}
\end{figure}

\begin{figure}[bth]
  \includegraphics[width=.30\textwidth]{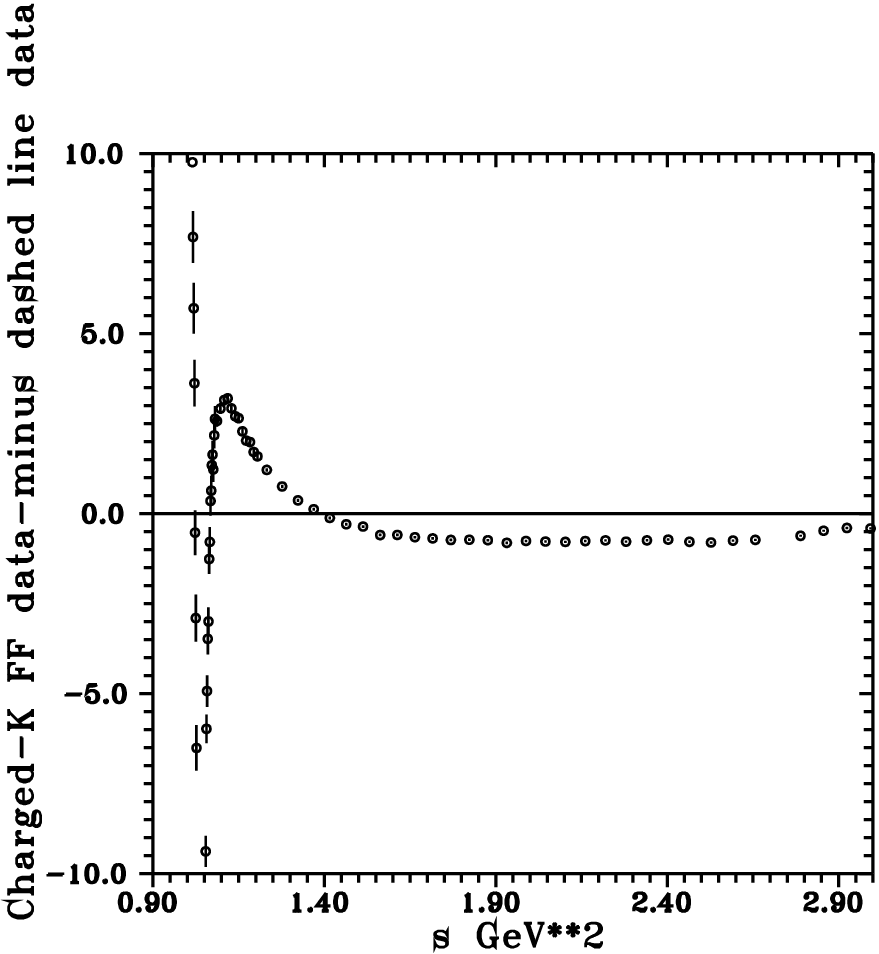}\\
  \caption{Damped oscillation structures revealed by a subtraction of dashed line data in Fig.4 from selected $|F_{K^{\pm}}(s)|$ data with errors.}\label{FIG. 5:}
\end{figure}

   Next we will show that when charged K-meson data are accurately described by some proper physically well founded model of the K-meson EM structure, no damped
oscillatory structures are observed.

   We will use the model that has been suggested in the paper \cite{DD} as the universal Unitary and Analytic (U$\&$A) EM structure model of hadrons and which unifies
three aspects:\\
1.) experimental fact of a creation of unstable vector-meson resonances (see \cite{PDG}), mainly identified in the electron-positron annihilation
processes into hadrons,\\
2.) two square root branch cut approximation of the analytic properties of FFs in the complex plane of c.m. energy squared $s$,\\
3.) and the correct asymptotic behavior of FFs as predicted by the quark model of hadrons.

   To describe the selected data in Fig.2 in the framework of the U$\&$A approach \cite{DD} we begin by splitting of the charged
K-meson FF, due to special transformation properties of the K-meson EM current in the isospin space, into a sum of the isoscalar and isovector parts
\begin{eqnarray}\label{KFFv}
 F_{K^{\pm}}(s)=F^s_K(s)+F^v_K(s)
\end{eqnarray}
with their norms
\begin{eqnarray}\label{Knorm}
F^s_K(0)=F^v_K(0)=\frac{1}{2}.
\end{eqnarray}

   Next we must deal with the question of what isoscalar and isovector resonances experimentally confirmed in \cite{PDG} will saturate the isoscalar and
isovector parts of the charged kaon FF, because in practice one can not consider all existing resonances with isospin $I=0$ and isospin $I=1$, as they will produce all together 27 free parameters, too much to be evaluated in the analysis of up-to-date existing experimental information. As a result a selection of contributing resonances has to be carried out and is done in the following way.

   The K-mesons include also strange quarks, therefore one could expect that in a description of the selected data in Fig.2 the $\phi$ resonances with the
isospin value $I=0$ will be dominant and their masses and widths will be left as free parameters of the model. There is no doubt about the inclusion of the $\phi(1020)$ resonance clearly seen in Fig.2. From Fig.2 (in more detail in Fig.3) between 2.0 GeV$^2$ and 7.0 GeV$^2$ one finds two bumps corresponding approximately to $\phi'(1680)$ and $\phi''(2170)$. The first bump could contain a contribution also from another resonance with $I=0$, $\omega''(1650)$, which is therefore not excluded. There is no indication in existing data of a contribution of the $\omega'(1420)$, therefore we don't consider it in our analysis.

   On the other hand we have an experience that one can not achieve satisfactory description of existing data without inclusion of the ground state resonances
$\rho(770)$ and $\omega(782)$. Further, because contributions of the isovector part of the K-meson FF, though not dominant, can not be ignored, we include contributions of all three $\rho$-mesons, however with fixed parameters from the paper \cite{BDLDK}, in which one can find reasons why not to utilize their parameters from \cite{PDG}.

    Masses and widths of $\omega(782)$ and $\omega''(1650)$ are also fixed at the PDG values.

   Then the U$\&$A model of the K-meson EM structure takes the following form.

   The isoscalar FF with 5 experimentally confirmed \cite{PDG} isoscalar resonances is
\begin{small}
\begin{eqnarray}\label{Fissc}
 F^s_K[V(s)]=(\frac{1-V^2}{1-V^2_N})^2\nonumber\\
 \times[\sum_{s=\omega,\phi}\frac{(V_N-V_s)(V_N-V^*_s)(V_N-1/V_s)(V_N-1/V^*_s)}
 {(V-V_s)(V-V^*_s)(V-1/V_s)(V-1/V^*_s)}(\frac{f_{sKK}}{f_s})\\
 +\sum_{s=\phi',\omega'',\phi''}\frac{(V_N-V_s)(V_N-V^*_s)(V_N+V_s)(V_N+V^*_s)}
 {(V-V_s)(V-V^*_s)(V+V_s)(V+V^*_s)}(\frac{f_{sKK}}{f_s})].\nonumber
\end{eqnarray}
\end{small}
where the concrete form of individual terms depends on the numerical value of the effective inelastic threshold $s^s_{in}$ which is found numerically by the fit of the model to charged K-meson EM FF selected data.

   In the previous expression
\begin{small}
\begin{equation}\label{comftrs1}
   V(s)=i\frac{\sqrt{(\frac{s^s_{in}-s^s_0}{s^s_0})^{1/2}+(\frac{s-s^s_0}{s^s_0})^{1/2}}-
   \sqrt{(\frac{s^s_{in}-s^s_0}{s^s_0})^{1/2}-(\frac{s-s^s_0}{s^s_0})^{1/2}}}
   {\sqrt{(\frac{s^s_{in}-s^s_0}{s^s_0})^{1/2}+(\frac{s-s^s_0}{s^s_0})^{1/2}}+
   \sqrt{(\frac{s^s_{in}-s^s_0}{s^s_0})^{1/2}-(\frac{s-s^s_0}{s^s_0})^{1/2}}}
\end{equation}
\end{small}
is the conformal mapping of the four sheeted Riemann surface into one V-plane, and $V_N=V(0)$ is a normalization point in V-plane with $s^s_0=9m^2_\pi$.
\begin{small}

   The isovector FF with 3 experimentally confirmed \cite{PDG} isovector resonances
$\rho(770), \rho'(1450), \rho''(1700)$ takes the form
\begin{eqnarray}
 F^v_K[W(s)]=(\frac{1-W^2}{1-W^2_N})^2
 [\frac{(W_N-W_\rho)(W_N-W^*_\rho)(W_N-1/W_\rho)(W_N-1/W^*_\rho)}
 {(W-W_\rho)(W-W^*_\rho)(W-1/W_\rho)(W-1/W^*_\rho)}(\frac{f_{\rho\pi\pi}}{f_\rho})\\
 + \sum_{v=\rho',\rho''}\frac{(W_N-W_v)(W_N-W^*_v)(W_N+W_v)(W_N+W^*_v)}
 {(W-W_v)(W-W^*_v)(W+W_v)(W+W^*_v)}(\frac{f_{v\pi\pi}}{f_v})],\nonumber
\end{eqnarray}
\end{small}
and again the structure of individual terms depends on the value of the effective inelastic threshold $s^v_{in}$ numerically evaluated in the fitting procedure of the model to charged K-meson EM FF data.

   In the previous expression
\begin{small}
\begin{equation}\label{comftrs2}
   W(s)=i\frac{\sqrt{(\frac{s^v_{in}-s^v_0}{s^v_0})^{1/2}+(\frac{s-s^v_0}{s^v_0})^{1/2}}-
   \sqrt{(\frac{s^v_{in}-s^v_0}{s^v_0})^{1/2}-(\frac{s-s^v_0}{s^v_0})^{1/2}}}
   {\sqrt{(\frac{s^v_{in}-s^v_0}{s^v_0})^{1/2}+(\frac{s-s^v_0}{s^v_0})^{1/2}}+
   \sqrt{(\frac{s^v_{in}-s^v_0}{s^v_0})^{1/2}-(\frac{s-s^v_0}{s^v_0})^{1/2}}}
\end{equation}
\end{small}
is a conformal mapping of the four sheeted Riemann surface, on which $F^v_K(s)$ is defined, into one W-plane, and $W_N=W(0)$
is a normalization point in W-plane, with $s^v_0=4m^2_\pi$.

   As a result, the U$\&$A model of the K-meson EM structure depends altogether on 14 free parameters and their numerical values (see TABLE I) have been evaluated in the
analysis of selected data from Fig.2.
\begin{table}\label{TABLE I}
\caption{Parameter values of the analysis of selected data on the $|F_{K^{\pm}}(s)|$ with minimum of $\chi^2/ndf=1.79$. \label{TABLE I}}
\begin{tabular}{c}
\hline
$s^s_{in}=(1.0984 \pm 0.2292)$ [GeV$^2$];\\
$m_{\phi}=(1019.298 \pm 0.063)$ [MeV]; $\Gamma_{\phi}= (4.304 \pm 0.083)$ [MeV];  $(f_{\phi K K}/f_\phi)=0.331 \pm 0.063$; \\
$m_{\phi'}=(1656.620 \pm 4.969)$ [MeV]; $\Gamma_{\phi'}= 356.860 \pm 4.444)$ [MeV]; $(f_{\phi' K K}/f_{\phi'})=-.568 \pm 0.102$; \\
$m_{\phi''}=(2001.300 \pm 22.817)$ [MeV]; $\Gamma_{\phi''}= (530.502 \pm 34.300)$ [MeV];\\
$(f_{\phi'' K K}/f_{\phi''})= 1/2- (f_{\omega K K}/f_\omega)-(f_{\omega'' K K}/f_{\omega''})-(f_{\phi K K}/f_\phi)-(f_{\phi' K K}/f_{\phi'})$; \\
$(f_{\omega K K}/f_\omega)=0.273 \pm 0.044$; $(f_{\omega'' K K}/f_{\omega''})=0.354 \pm 0.103$;\\
$s^v_{in}=(1.6765 \pm 0.1337)$ [GeV$^2$];\\
$(f_{\rho' K K}/f_{\rho'})=1/2-(f_{\rho K K}/f_\rho)-(f_{\rho'' K K}/f_{\rho''})$\\
$(f_{\rho K K}/f_\rho)=0.440 \pm 0.017$; $(f_{\rho'' K K}/f_{\rho''})=0.036 \pm 0.005$;\\
\hline
\end{tabular}
\end{table}

    The corresponding accurate description of these charged K-meson EM FF data is presented in Fig.6 by full line. One finds the description of the resonant region
between 2.0 GeV$^2$ and 7.0 GeV$^2$ in more detail in Fig.7.
\begin{figure}
    \includegraphics[width=0.30\textwidth]{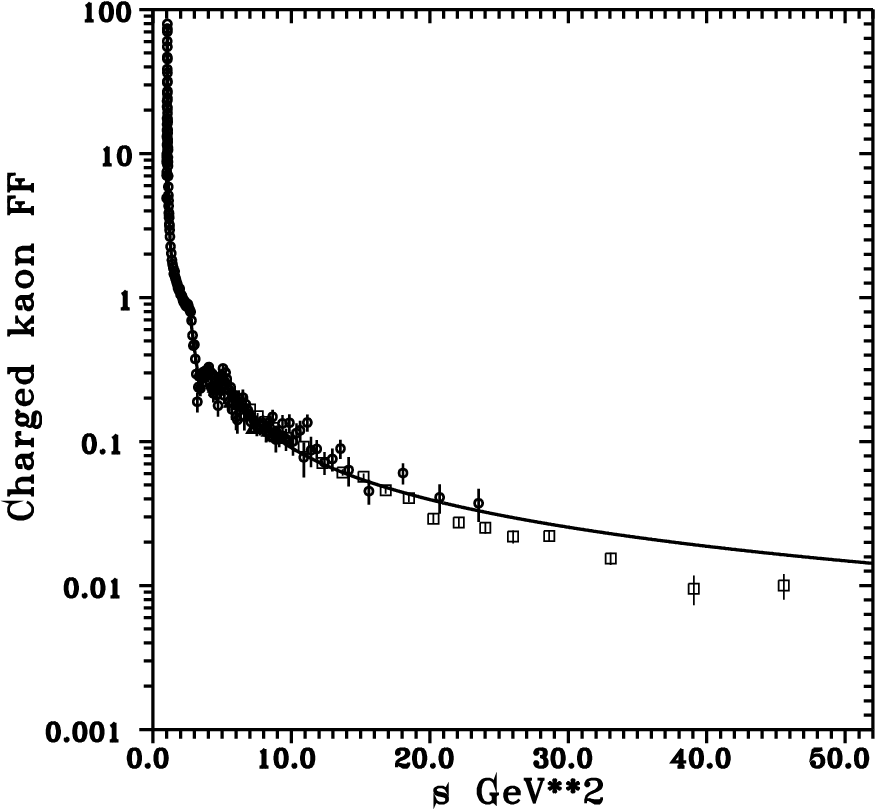}\hspace{0.3cm}
    \includegraphics[width=0.30\textwidth]{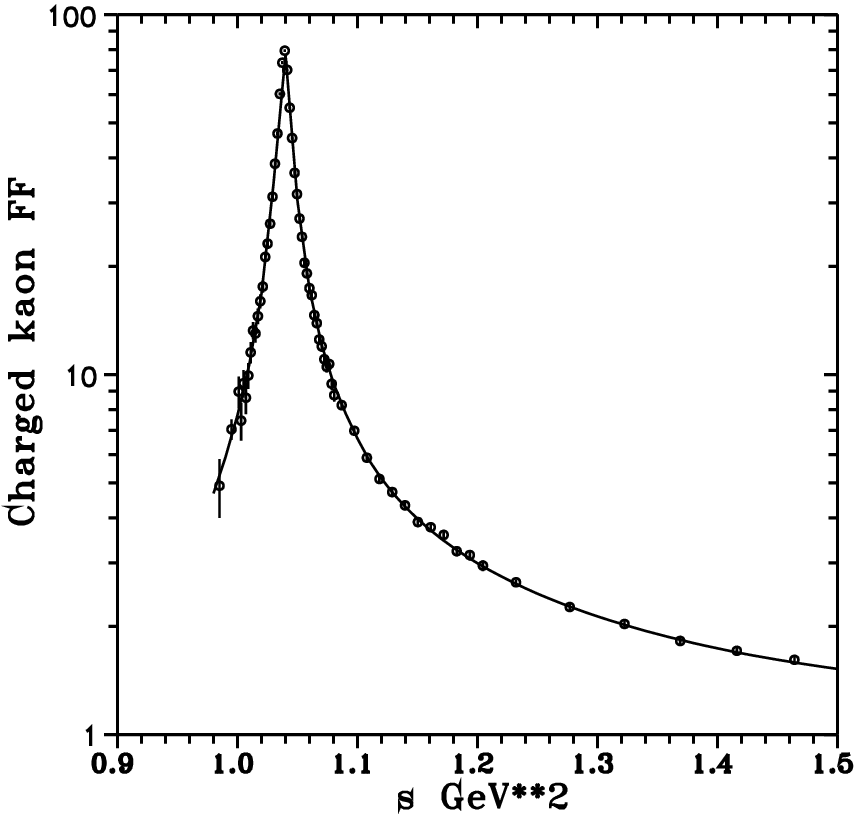}\\
\caption{Charged kaon EM FF data described by U$\&$A model}\label{Fig.6}
\end{figure}
\begin{figure}[bth]
  \includegraphics[width=.30\textwidth]{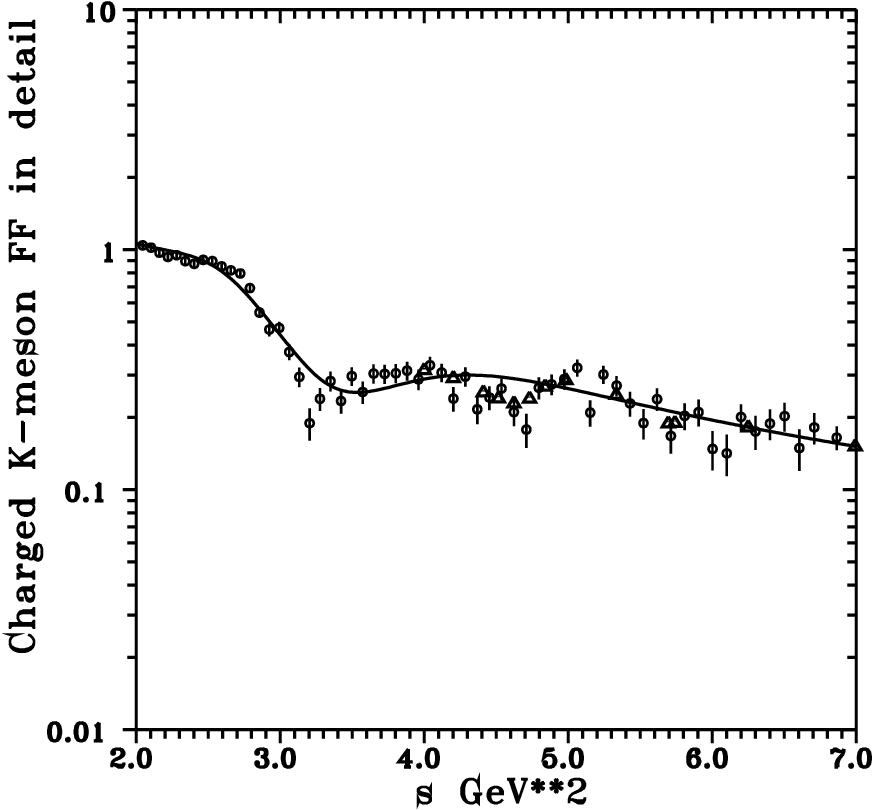}\\
  \caption{Description of charge K-meson EM FF data by U$\&$A model between 2.0 GeV$^2$ and 7.0 GeV$^2$.}\label{Fig.7}
\end{figure}
\begin{figure}[bth]
  \includegraphics[width=.30\textwidth]{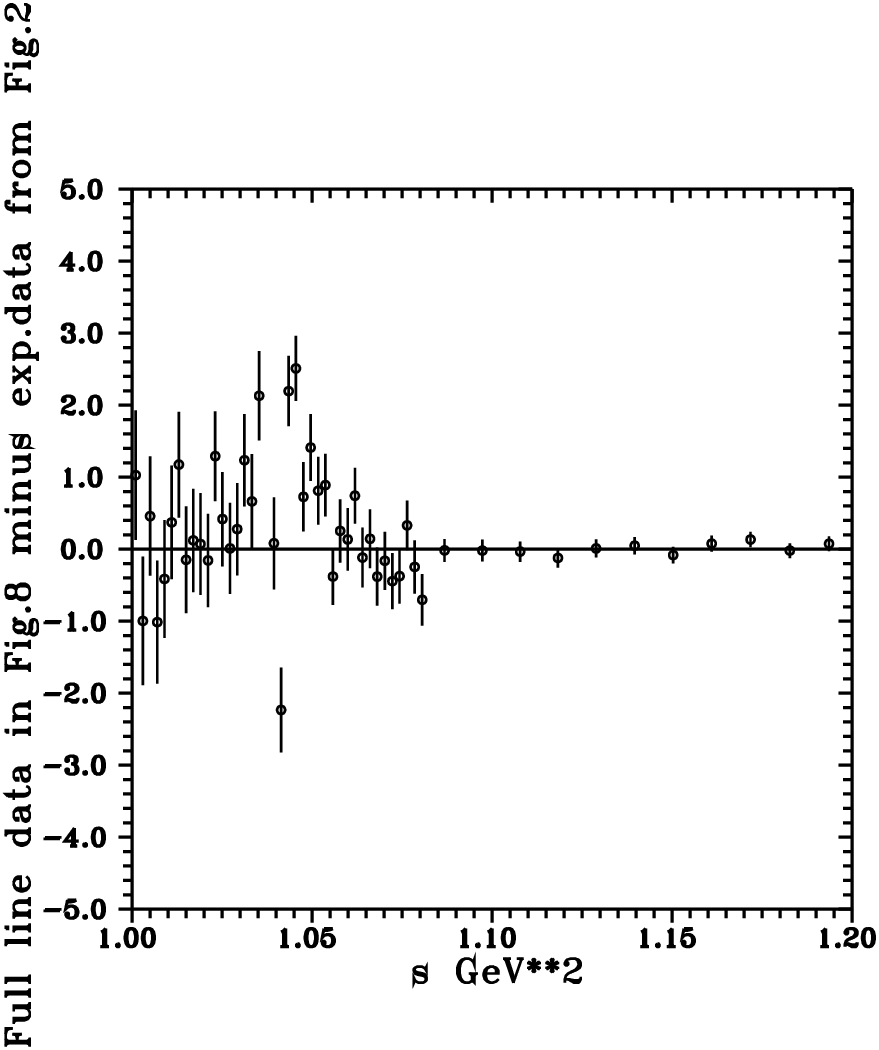}\\
  \caption{No damped oscillatory structures appear if $|F_{K^{\pm}}(s)|$ data are accurately described by U$\&$A model of K-meson EM structure as presented in Fig.6.}\label{Fig.8}
\end{figure}

   Lastly, if the full line in Fig.6 is subtracted from selected charged K-meson FF data in Fig.2 with errors,
no damped oscillating structures are observed around the line crossing the zero as it is shown in Fig.8.

\section{INVESTIGATION OF DAMPED OSCILLATING STRUCTURES FROM NEUTRAL K-MESON EM FF TIMELIKE DATA}

   The timelike data on the neutral K-meson EM FF $|F_{K^0}(s)|$, which completely describes the neutral K-meson EM structure, can be obtained from the measured total
cross section $\sigma^{bare}_{tot}(e^+e^- \to K_S K_L)$, which is, however, more difficult to measure than $\sigma^{bare}_{tot}(e^+e^- \to K^+ K^-)$.
As a result one can expect also that the obtained experimental information on $|F_{K^0}(s)|$ will be of lower quality than the data on $|F_K^{\pm}(s)|$.

   According to our knowledge there are no data on the function $|F_{K^0}(s)|$ with errors published during the last decade, nevertheless we calculate them in this paper
by means of the relation
\begin{eqnarray}\label{nKFF}
  |F_{K^0(s)}|=\sqrt{\sigma^{bare}_{tot}(e^+e^- \to K_S K_L)\frac{3s}{\pi\alpha^2\beta^3_{K^0}}}
\end{eqnarray}
with $\beta_{K^0}(s)=\sqrt{1-\frac{4 m^2_{K^0}}{s}}$, $\alpha$=1/137, from one recent measurement \cite{Lees5} of the process $e^+e^- \to K_SK_L(\gamma)$ by the ISR technique at the interval of energy values $s$ (1.1664-4.84) GeV$^2$ and from two measurements \cite{Kozyrev}, \cite{Ablikim7} by the scan method, the first one in the $\phi$-resonance region (1.0080-1.1236) GeV$^2$ and the second at the range of energies (4.0000-9.4864) GeV$^2$. This second measurement is the first measurement that has probed this interval of energies. Their numerical values with errors are given in Table II and graphically presented in Fig.9.
\begin{table}
	\caption{The data on $|F_{K^0}(s)|$ with errors. \label{TABLE II}}
	\newcolumntype{C}{>{\centering\arraybackslash}X}
	\begin{tabularx}{\textwidth}{cccccccc}
	\hline\\
	$s$ [GeV$^{2}$] & $|F_{K^0}(s)|\pm err.$ & $s$ [GeV$^{2}$] & $|F_{K^0}(s)|\pm err.$ & $s$ [GeV$^{2}$] & $|F_{K^0}(s)|\pm err.$ & $s$ [GeV$^{2}$] & $|F_{K^0}(s)|\pm err.$\\
	\hline
	$1.0080$ &  $11.4683 \pm 0.3506$ & $1.0588$ &  $16.7599 \pm 0.1729$ &  $1.9600$ & $0.2369 \pm 0.0592$ & $4.2025$  & $0.1120 \pm 0.0496$ \\
	$1.0201$ &  $19.6100 \pm 0.1093$ & $1.0692$ &  $10.9758 \pm 0.1380$ &  $2.0736$ & $0.2821 \pm 0.0529$ & $4.4100$  & $0.0829 \pm 0.0004$ \\
	$1.0241$ &  $27.1841 \pm 0.1405$ & $1.0816$ &  $07.8863 \pm 0.1274$ &  $2.1904$ & $0.2968 \pm 0.0453$ & $4.4944$  & $0.0713 \pm 0.0035$ \\
	$1.0302$ &  $36.4652 \pm 0.4169$ & $1.1025$ &  $05.0943 \pm 0.0752$ &  $2.3104$ & $0.3999 \pm 0.0394$ & $4.6225$  & $0.0653 \pm 0.0136$ \\
	$1.0323$ &  $45.4557 \pm 0.2066$ & $1.1046$ &  $05.1158 \pm 0.1376$ &  $2.4336$ & $0.4022 \pm 0.0390$ & $4.7306$  & $0.0748 \pm 0.0075$ \\
	$1.0343$ &  $57.6351 \pm 0.1353$ & $1.1236$ &  $03.8750 \pm 0.1138$ &  $2.5600$ & $0.4505 \pm 0.0322$ & $4.8400$  & $0.0738 \pm 0.0057$ \\
    $1.0363$ &  $70.4012 \pm 0.1847$ & $1.1664$ &  $02.4831 \pm 0.0778$ &  $2.6896$ & $0.4488 \pm 0.0351$ & $4.9836$  & $0.0876 \pm 0.0666$ \\
	$1.0384$ &  $81.3097 \pm 0.1731$ & $1.2544$ &  $01.2826 \pm 0.0687$ &  $2.8224$ & $0.3531 \pm 0.0381$ & $5.3333$  & $0.0716 \pm 0.0048$ \\
	$1.0404$ &  $77.0351 \pm 0.1361$ & $1.3456$ &  $00.8939 \pm 0.0603$ &  $2.9584$ & $0.2534 \pm 0.0536$ & $5.6949$  & $0.0342 \pm 0.0055$ \\
	$1.0424$ &  $59.0975 \pm 0.1734$ & $1.4400$ &  $00.7436 \pm 0.0578$ &  $3.0976$ & $0.1659 \pm 0.0528$ & $5.7408$  & $0.0367 \pm 0.0038$ \\
	$1.0445$ &  $48.1169 \pm 0.1663$ & $1.5376$ &  $00.5341 \pm 0.0553$ &  $3.2400$ & $0.0872 \pm 0.0727$ & $6.9929$  & $0.0437 \pm 0.0050$ \\
	$1.0465$ &  $37.7921 \pm 0.2656$ & $1.6384$ &  $00.4013 \pm 0.0520$ &  $3.3856$ & $0.1015 \pm 0.0508$ & $7.0034$  & $0.0455 \pm 0.0043$ \\
	$1.0506$ &  $26.1639 \pm 0.3264$ & $1.7424$ &  $00.3877 \pm 0.0538$ &  $3.5344$ & $0.0724 \pm 0.0724$ & $8.4100$  & $0.0253 \pm 0.0036$ \\
	$1.0568$ &  $18.6920 \pm 0.1371$ & $1.8496$ &  $00.3253 \pm 0.0529$ &  $4.0000$ & $0.1223 \pm 0.0663$ & $9.4864$  & $0.0257 \pm 0.0030$ \\	
\hline
\end{tabularx}
\end{table}

\begin{figure}
    \includegraphics[width=0.30\textwidth]{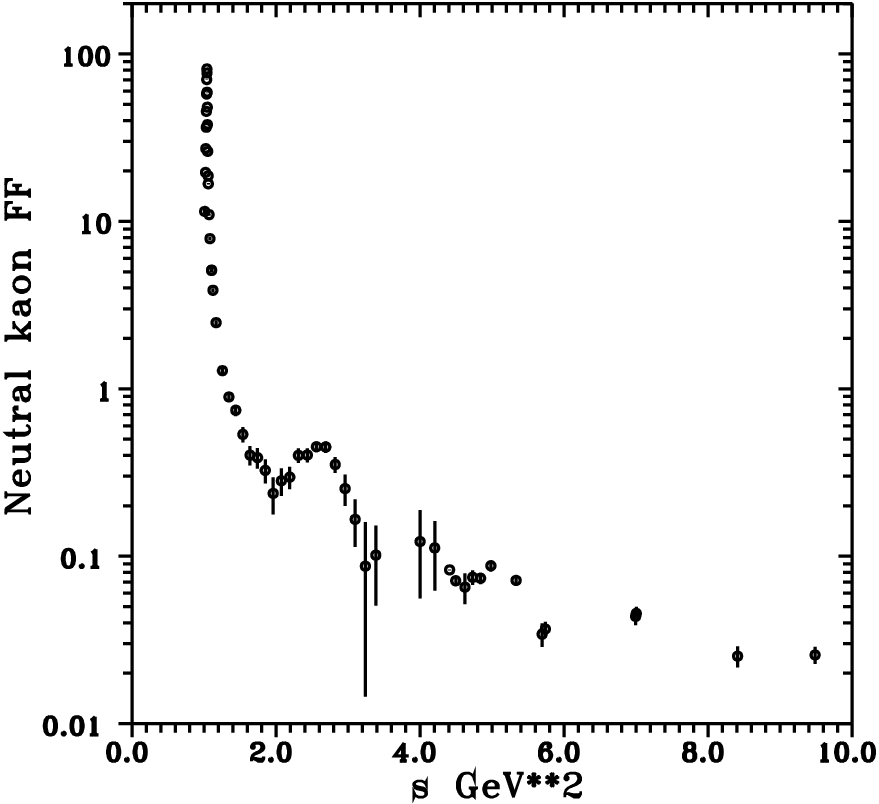}\hspace{0.3cm}
    \includegraphics[width=0.30\textwidth]{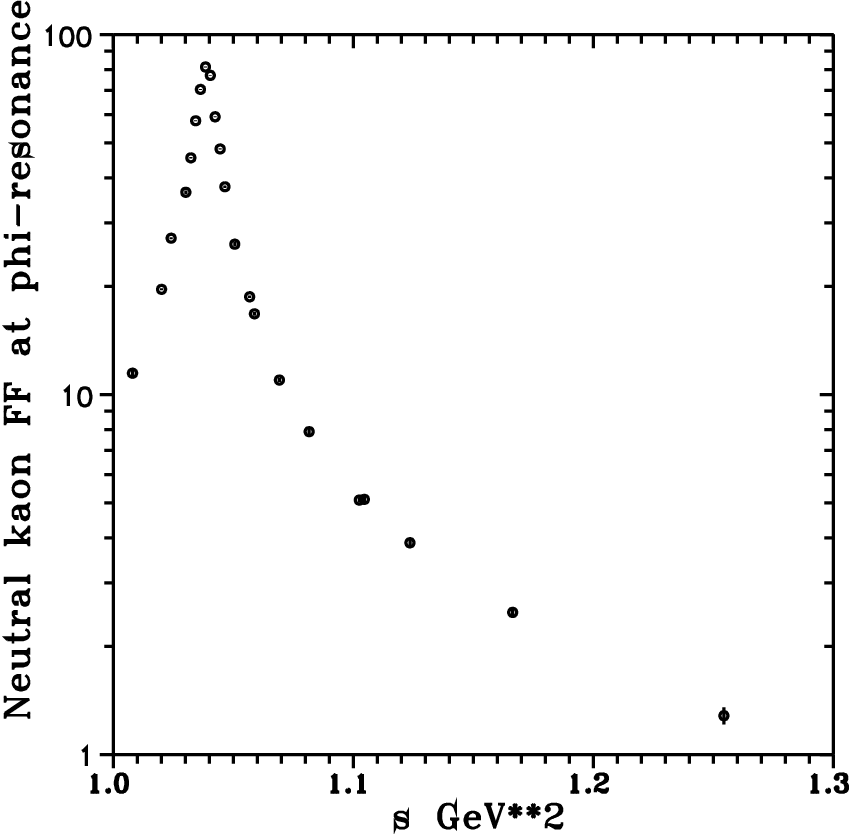}\\
\caption{Neutral kaon EM FF data.}\label{Fig.9}
\end{figure}

   In order to study eventual damped oscillatory structures from the neutral K-meson EM FF timelike data by using the same procedure as for the proton, again the
modified formula (\ref{fortfunct}) with the third parameter 0.71 GeV$^2$ to be a free parameter A3 is utilized in a fitting procedure of the data in Fig.9. Their best description has been achieved with parameter values A3=$0.9269\pm0.0003$ GeV$^2$, $m^2_a$=$-.9267\pm0.0016$ GeV$^2$, A=$0.4729\pm0.0061$ and is graphically presented in Fig.10 by the dashed line.

\begin{figure}
    \includegraphics[width=0.30\textwidth]{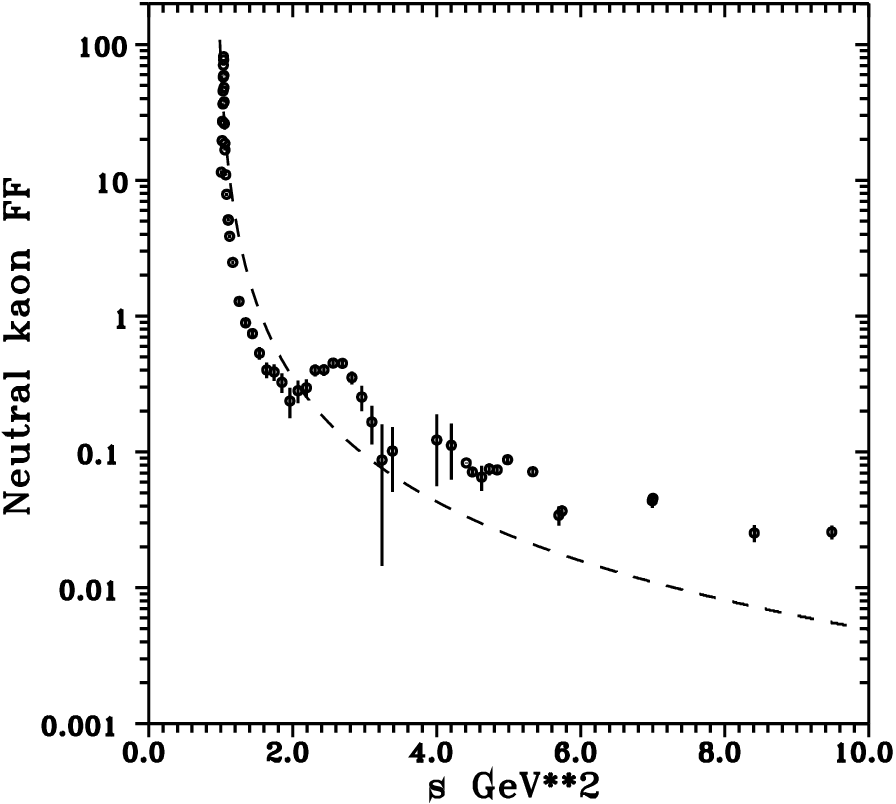}\hspace{0.3cm}
    \includegraphics[width=0.30\textwidth]{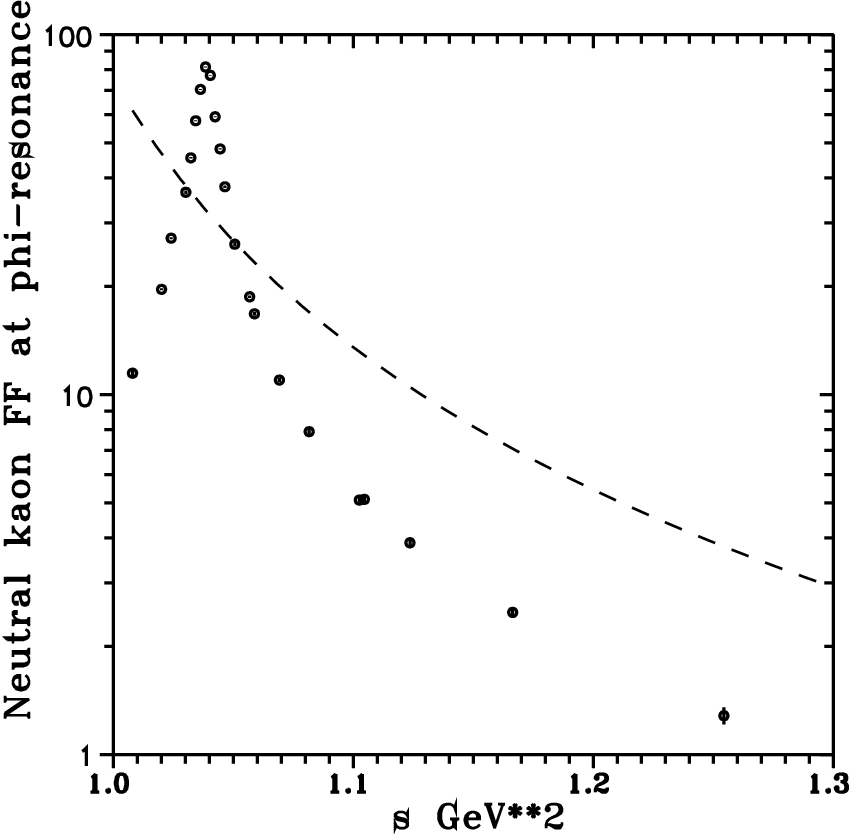}\\
\caption{Neutral kaon EM FF data optimally described by dashed line as generated by the modified three parametric formula (\ref{fortfunct})}\label{Fig.10}
\end{figure}

   If the dashed line is subtracted from the neutral K-meson FF data with errors in Fig.9, damped oscillating structures can be observed around the line crossing the
zero as depicted in Fig.11.
\begin{figure}[bth]
  \includegraphics[width=.30\textwidth]{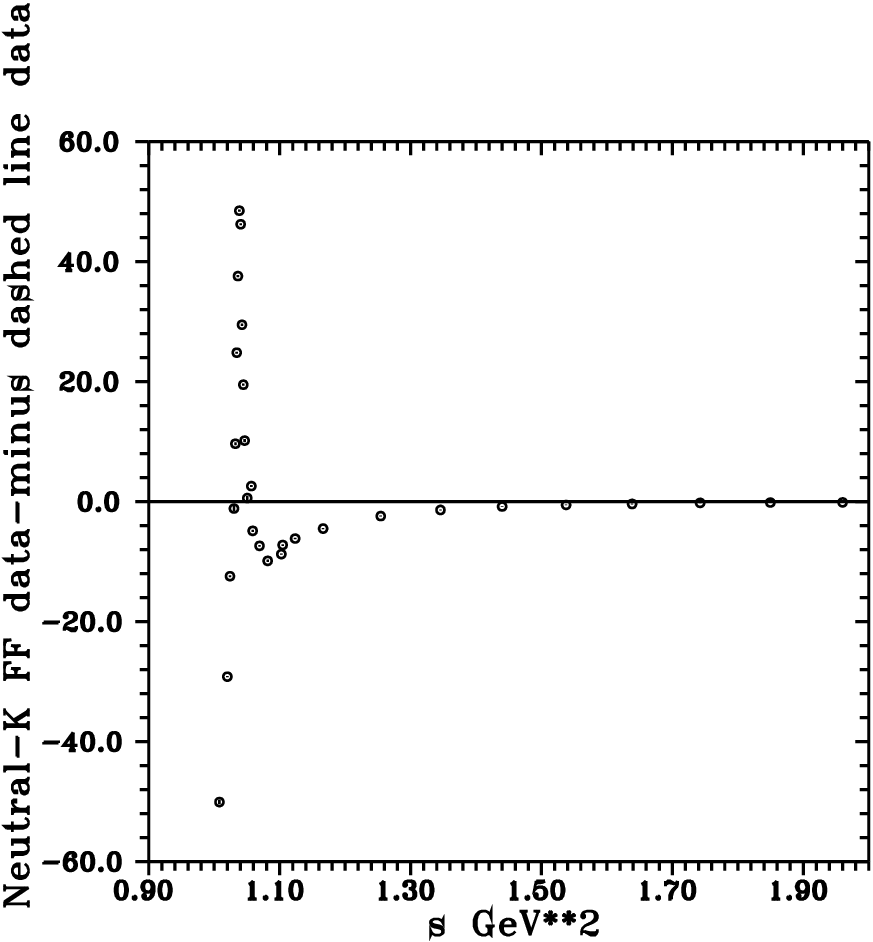}\\
  \caption{Damped oscillation structures revealed by a subtraction of dashed line data in Fig.10 from  $|F_{K^0}(s)|$ data with errors given in TABLE II.}\label{FIG. 11:}
\end{figure}

   Comparing the oscillations depicted in the Fig.11 with those of the Fig.5 one can see that there is an analogy between the behaviors of the damped oscillations in
the K-meson FF data and those of damped oscillations in the nucleon "effective" FF data, as presented in Fig.1 and in Fig.11b of \cite{Ablikim8}. In both cases the oscillations in the data for one particle (say the charged K-meson) correspond roughly to the oscillations in the data for its partner particle in the respective isospin doublet (say the neutral K-meson) if reflected through the axis y=0. Where the oscillations for charged kaons have maxima the oscillations for neutral kaons have minima, and vice versa. The same property can be observed in the oscillations of proton and neutron data.

   This feature seems to be interesting and it has to be resolved by some serious physical arguments.

   Now, as in the case of the charged K-meson EM FF data, we shall try to demonstrate that if neutral K-meson EM FF data in Fig.9 are accurately described by a proper
physically well founded model, no damped oscillatory structures is observed.

   The simplest way to obtain such a description is to exploit the special transformation properties of the K-meson EM current in the isospin space, from which it
directly follows that we can write the neutral K-meson EM FF as the difference of the isoscalar and isovector parts
\begin{eqnarray}\label{KFFs}
 F_{K^0}(s)=F^s_K(s)-F^v_K(s)
\end{eqnarray}
where $F^s_K(s)$ and $F^v_K(s)$ are the same as those in (\ref{KFFv}). The charged and neutral K-meson EM FFs then depend on the same set of parameters of the $U\&A$ model.

   Thus a substitution of the numerical values of parameters from TABLE I into the K-meson EM FF U$\&$A model (\ref{Fissc})-(\ref{comftrs2}) should lead through
(\ref{KFFs}) to an accurate description of data in TABLE II.

\begin{figure}[bth]
    \includegraphics[width=0.30\textwidth]{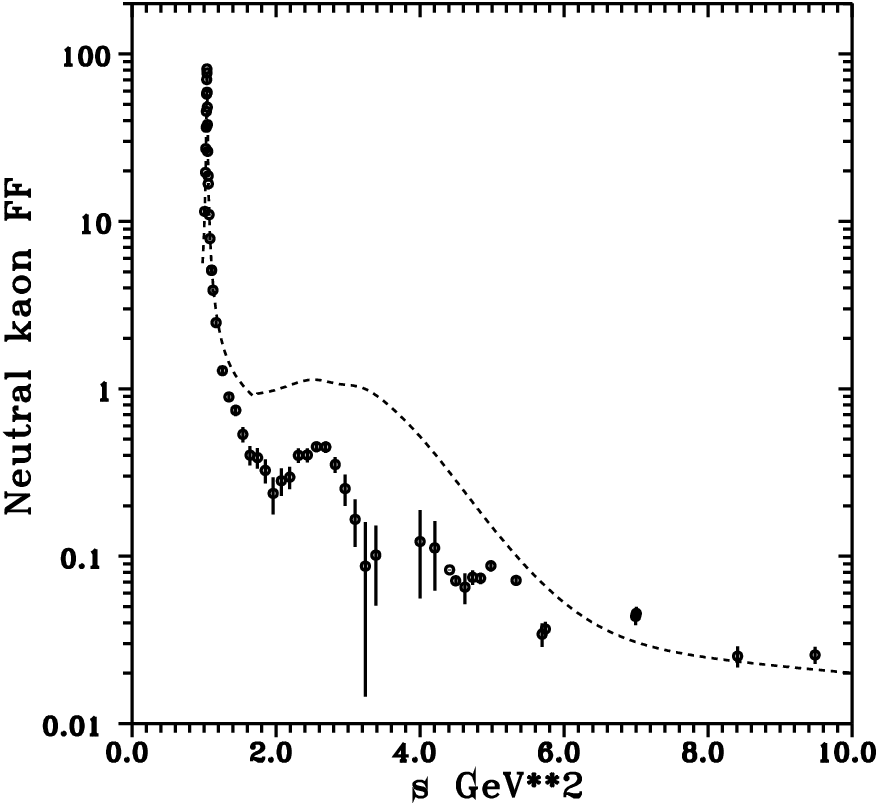}\hspace{0.3cm}
    \includegraphics[width=0.30\textwidth]{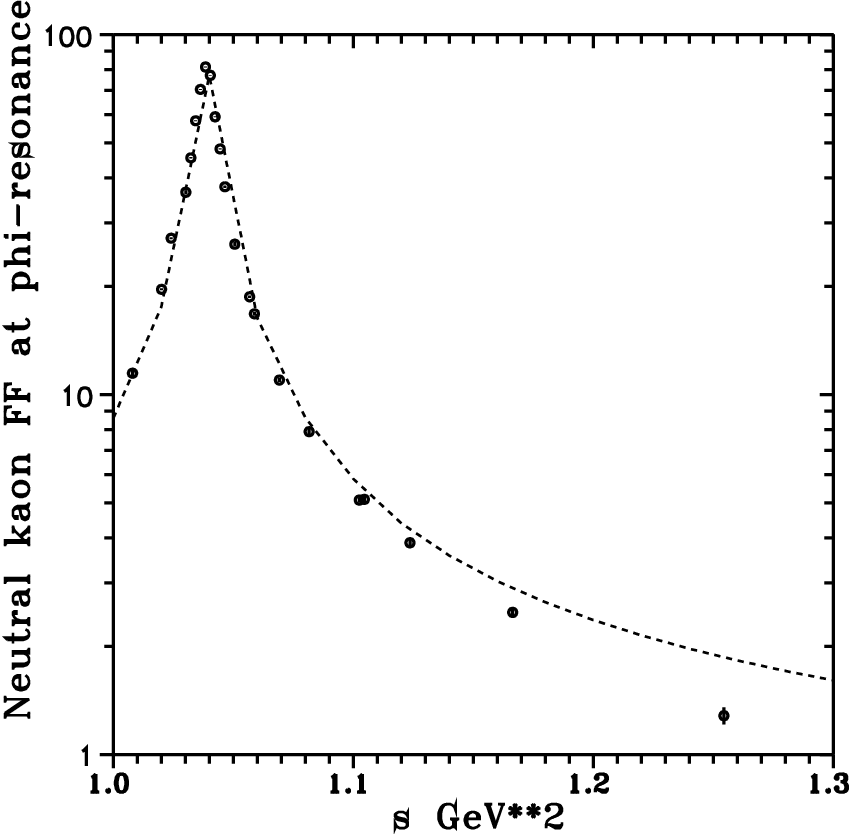}\\
\caption{Predicted behavior of $|F_{K^0}(s)|$ from data on charged K-meson FF by exploiting special transformation properties of the kaon EM current in the isospin space does not describe experimental data on it accurately.}\label{Fig.12}
\end{figure}

   However, as can be explicitly seen in Fig.12, especially at the energy region from 1.4 GeV$^2$ up to 6.0 GeV$^2$, the predicted dotted line does not follow the data
for $|F_{K^0}(s)|$ accurately.

   There are several ways how to interpret this discrepancy. It can be a consequence of the non-conservation of isospin symmetry or some subtle error in the definition
of the isoscalar and isovector parts in the $U\&A$ model. Furthermore, it is possible that the data are not precise enough to allow for a good separation of the isoscalar and isovector  components by a fitting procedure on charged kaon data only (as opposed to fitting on both the charged and neutral kaon data together). And lastly, it could be that the two experimental data sets are simply inconsistent. Nevertheless, resolving this issue is not important for the purpose of the current article.

   So, in order to achieve an accurate description of data in Fig.9 and to look for damped oscillation structures from the neutral EM FF timelike data, one has to carry
out a direct fitting procedure of the latter data by the K-meson EM FF $U\&A$ model (\ref{Fissc})-(\ref{comftrs2}). This yields the 14 parameters whose numerical values are presented in TABLE III. These parameters differ from parameter values in TABLE I. This may indicate that the data on charged K-meson EM FF and the data on neutral K-meson EM FF data are really in disagreement.

   The most accurate description of the $|F_{K^0}(s)|$ data with parameters in TABLE III is graphically presented in Fig.13 by the full line.

\begin{table}\label{TABLE III}
\caption{Parameter values from the analysis of selected data on the $|F_{K^0}(s)|$ by the K-meson EM FF U$\&$A
model (\ref{Fissc})-(\ref{comftrs2}) with minimum of $\chi^2/ndf=4.37$. \label{TABLE III}}
\begin{tabular}{c}
\hline
$s^s_{in}=(2.0728 \pm 0.0196)$ [GeV$^2$];\\
$m_{\phi}=(1019.158 \pm 0.176)$ [MeV]; $\Gamma_{\phi}= (4.214 \pm 0.030)$ [MeV];  $(f_{\phi K K}/f_\phi)=0.336 \pm 0.001$; \\
$m_{\phi'}=(1649.760 \pm 5.356)$ [MeV]; $\Gamma_{\phi'}= 340.032 \pm 12.438)$ [MeV]; $(f_{\phi' K K}/f_{\phi'})=-.217 \pm 0.001$; \\
$m_{\phi''}=(2028.740 \pm 46.556)$ [MeV]; $\Gamma_{\phi''}= (383.418 \pm 45.489)$ [MeV];\\
$(f_{\phi'' K K}/f_{\phi''})= 1/2- (f_{\omega K K}/f_\omega)-(f_{\omega'' K K}/f_{\omega''})-(f_{\phi K K}/f_\phi)-(f_{\phi' K K}/f_{\phi'})$; \\
$(f_{\omega K K}/f_\omega)=0.278 \pm 0.001$; $(f_{\omega'' K K}/f_{\omega''})=0.088 \pm 0.001$;\\
$s^v_{in}=(2.0077 \pm 0.0785)$ [GeV$^2$];\\
$(f_{\rho' K K}/f_{\rho'})=1/2-(f_{\rho K K}/f_\rho)-(f_{\rho'' K K}/f_{\rho''})$\\
$(f_{\rho K K}/f_\rho)=0.606 \pm 0.004$; $(f_{\rho'' K K}/f_{\rho''})=-.044 \pm 0.010$;\\
\hline
\end{tabular}
\end{table}

\begin{figure}
    \includegraphics[width=0.30\textwidth]{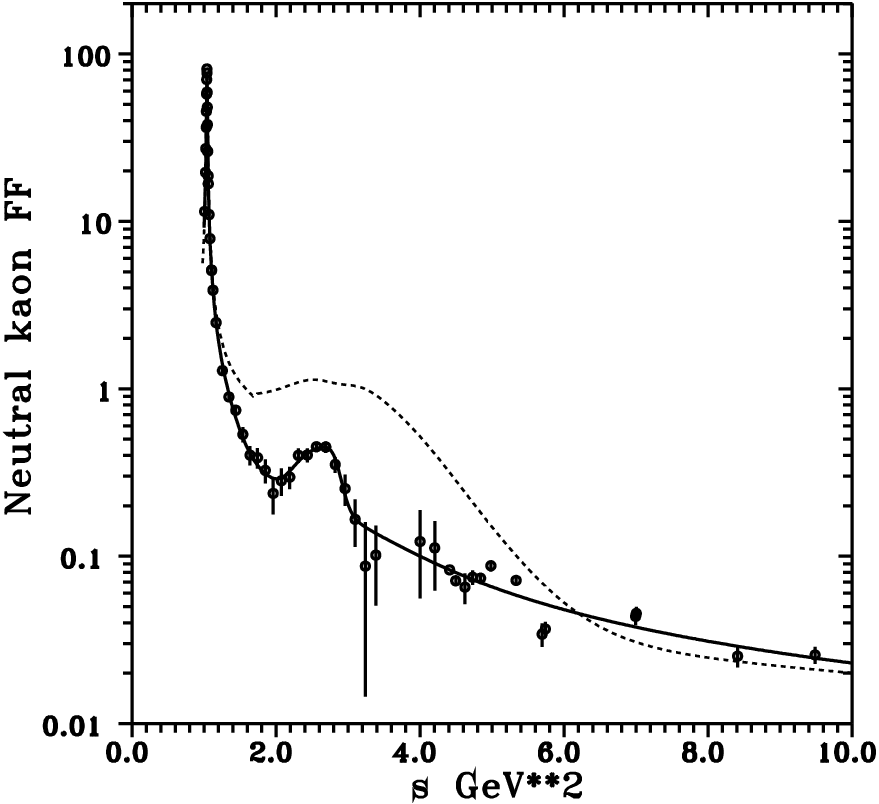}\hspace{0.3cm}
    \includegraphics[width=0.30\textwidth]{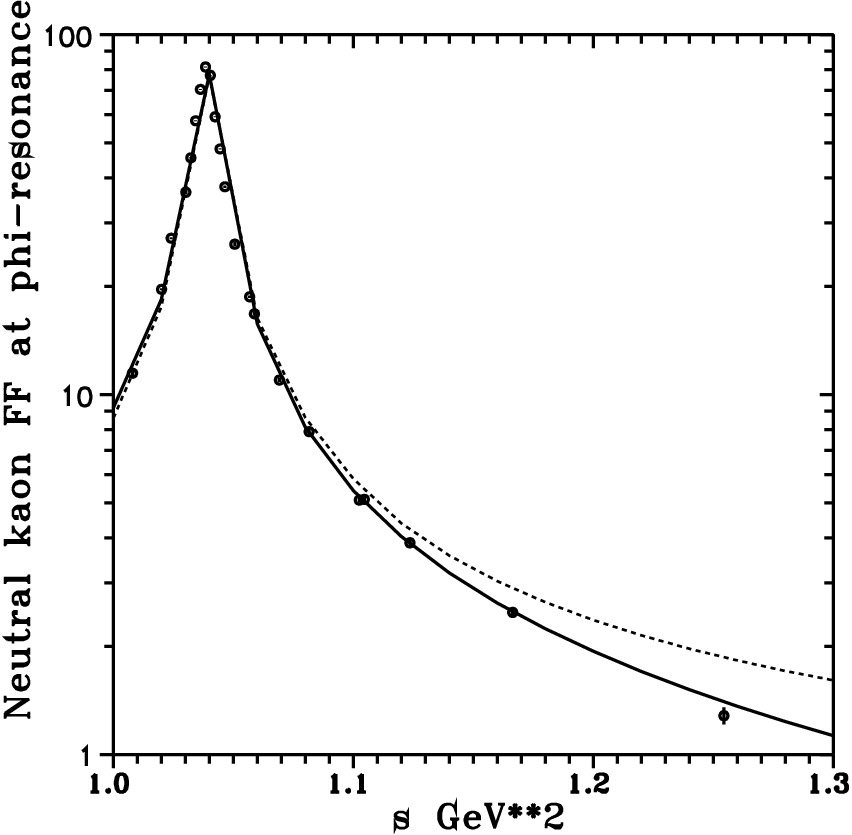}\\
\caption{Neutral kaon EM FF data described by the full line to be obtained by K-meson EM FF U$\&$A model (\ref{Fissc})-(\ref{comftrs2}) with parameters
of TABLE III}\label{Fig.13}
\end{figure}
\begin{figure}[bth]
  \includegraphics[width=.30\textwidth]{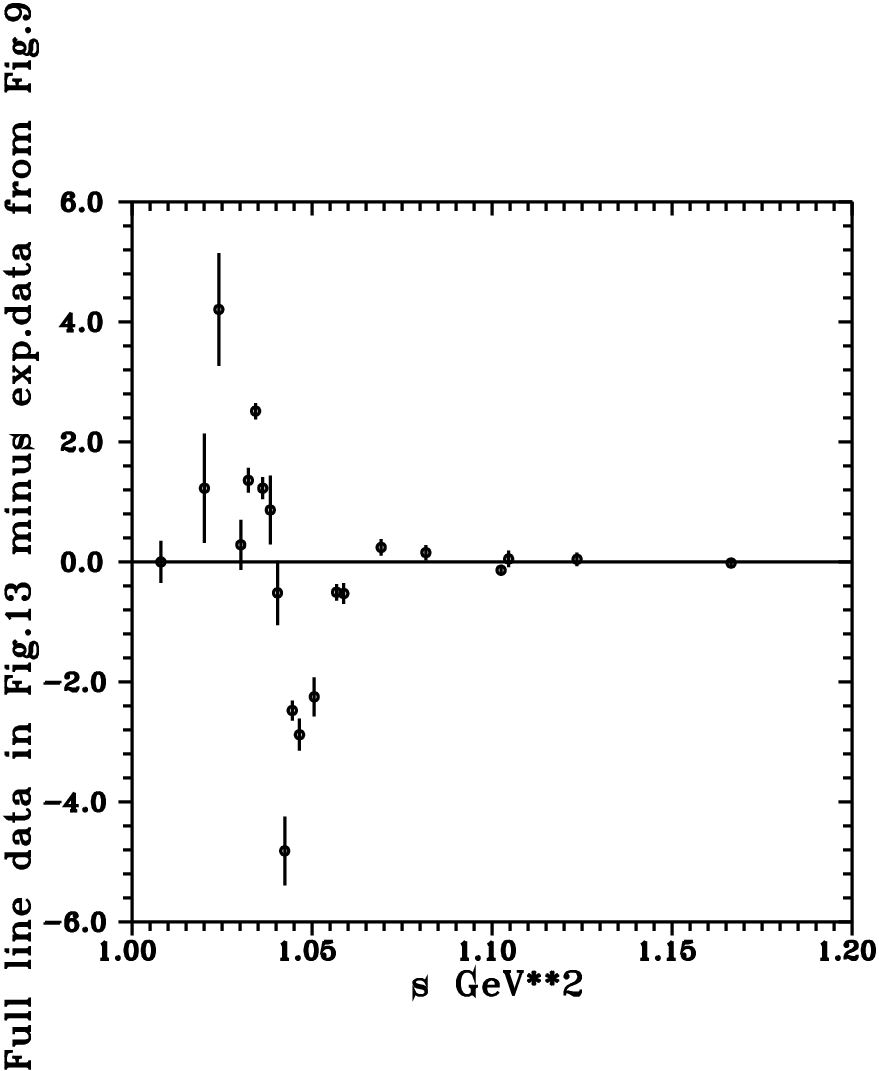}\\
  \caption{Points with errors obtained by a subtraction of full line data in Fig.13 from  $|F_{K^0}(s)|$ data with errors in Fig.9}\label{FIG. 14:}
\end{figure}

   If full line data in Fig.13 are subtracted from selected neutral K-meson EM FF data in TABLE II with errors, one obtains
points with errors around the line crossing the zero in Fig.14, from which, however, one can not explicitly declare, especially at the region of the $\phi$-resonance peak, that appear or do not appear damped oscillatory structures. In order to obtain a more decisive resolution more precise data on $|F_{K^0}(s)|$ are needed.

   On this place, for a completeness, it is interesting to investigate also the poorer neutral K-meson EM FF data, by exploiting special transformation properties of the
K-meson EM current in the isospin space leading to a splitting (\ref{KFFs}) of the charge K-meson EM FF into a sum of the isoscalar and isovector parts, to what extent they are able to reproduce existing more rich charged K-meson EM FF data.

   For this aim one substitutes the numerical values of parameters from TABLE III  into the K-meson EM FF $U\&A$ model (\ref{Fissc})-(\ref{comftrs2}),
together with (\ref{KFFv}), and compares calculated a dotted curve with $|F_{K^{\pm}}(s)|$ data as it is demonstrated in Fig.15.

\begin{figure}
    \includegraphics[width=0.30\textwidth]{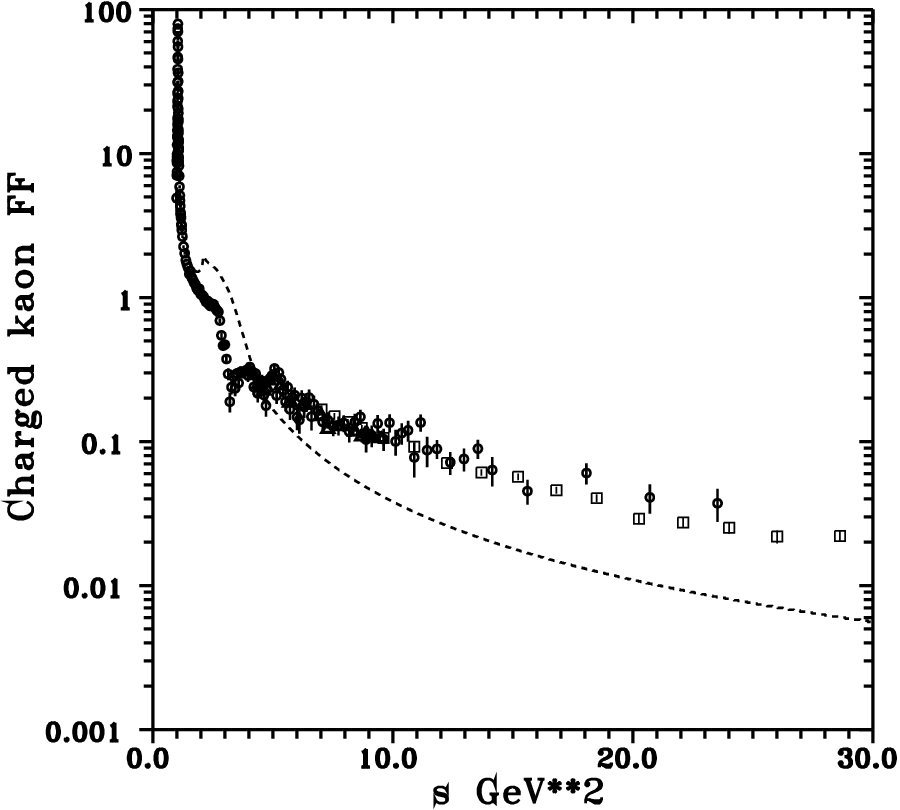}\hspace{0.3cm}
    \includegraphics[width=0.30\textwidth]{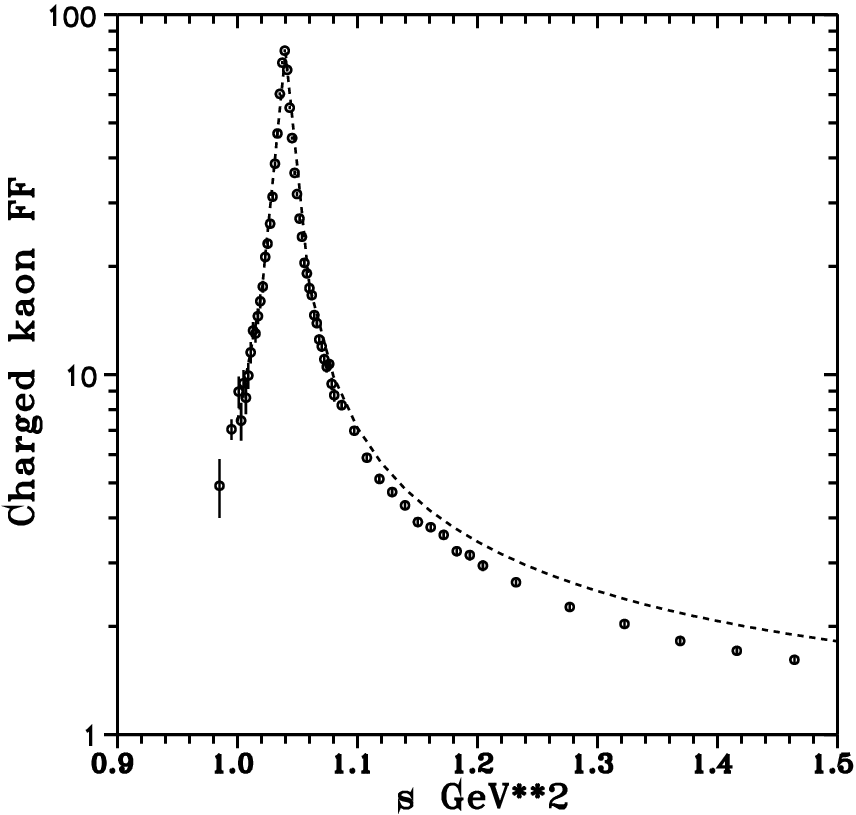}\\
\caption{Predicted behavior of $|F_{K^{\pm}}(s)|$ from data on neutral K-meson EM FF in TABLE II by exploiting special transformation properties of the kaon EM current in the isospin space does not describe experimental data on it at all, besides a little at the $\phi$ resonance region.}\label{Fig. 15}
\end{figure}

   A collation of Fig.12 with Fig.15 really indicate that the data on the charged and neutral K-meson EM FF may be probably in a disagreement.

\section{CONCLUSIONS AND DISCUSSION}

   The damped oscillating structures from the proton "effective" form factor data raised an interest to study damped oscillatory structures also from the
electromagnetic form factors data of other hadrons, for which solid data together with a physically well founded model for their accurate description exist.

   In this work the problem of the charged and neutral K-meson electromagnetic form factor data by using the same procedure as the one used in the case of the proton
in \cite{BianTom} has been investigated.

   The data on the charged kaon electromagnetic form factor are published together with $\sigma^{bare}_{tot}(e^+e^- \to K^+ K^-)$ data and
therefore in this paper they could be directly described in Figure 4 by the modified three parametric formula (\ref{fortfunct}) where instead of the magic nucleon number 0.71 additional free parameter A3 is introduced. If dashed line data in Figure 4 generated by this modified three parametric formula are subtracted from the charged K-meson electromagnetic form factor data in Figure 2, damped oscillation structures appear, as presented in Figure 5. If full line data in Figure 6 obtained by an accurate description of the charged K-meson electromagnetic form factor data by the physically well founded Unitary and Analytic model of the K-meson electromagnetic structure are subtracted from the charged K-meson electromagnetic form factor data in Figure 2 with errors, no damped oscillation structures appear (see Figure 8).

   The neutral K-meson electromagnetic form factor data can be extracted from $\sigma^{bare}_{tot}(e^+e^- \to K_S K_L)$ and first they were
calculated with errors using (\ref{nKFF}) (see TABLE II) and only then the modified formula, like in the charged case, has been applied for their description by the dashed line in Figure 10. If dashed line data generated by such three parametric formula are subtracted from the neutral K-meson electromagnetic form factor data in TABLE II, damped oscillation structures appear, as presented in Figure 11. If, however, full line data generated by the K-meson electromagnetic form factor Unitary and Analytic model (\ref{Fissc})-(\ref{comftrs2}) with parameter values in TABLE III in Figure 13 are subtracted from the neutral K-meson electromagnetic form factor data in TABLE II, one obtains points with errors around the line crossing the zero in Figure 14, from which, however, one can not explicitly declare, especially at the region of the $\phi$-resonance peak, wether the damped oscillatory structures are present. For the solution of this problem new, more precise data on  $|F_{K^0}(s)|$ are indispensable.

   Lastly the consistency of existing data on the charged and neutral K-meson electromagnetic form factor data has been investigated by exploiting special transformation
properties of the K-meson electromagnetic current in the isospin space, leading to a splitting of the charged and neutral K-meson electromagnetic for factors into a sum (\ref{KFFv}) and difference (\ref{KFFs}) of the same isoscalar and isovector parts, respectively. From the latter it follows, that both sets of experimental data on $|F_{K^{\pm}}(s)|$ and $|F_{K^0}(s)|$ should be described by the same numerical values of the parameters of the Unitary and Analytic model of the K-meson electromagnetic structure. However, evaluations of these parameters by means of the independent sets on $|F_{K^{\pm}}(s)|$ and $|F_{K^0}(s)|$ produce different values, as it is presented in TABLE I and TABLE III, respectively, which already indicate some inconsistency between both of them. To what extent, it is practically achieved by a prediction of
a behavior of $|F_{K^0}(s)|$ from the charged K-meson data and by a prediction of $|F_{K^{\pm}}(s)|$ from the neutral K-meson data, and their comparison with existing data in Figure 12 and Figure 15.

  Comparing Figure 12 with Figure 15 we have come to a conclusion that between charged and neutral K-meson electromagnetic form factor data some disagreement really
exists.

\medskip

   The authors of the paper would like to thank Erik Barto\v s for many useful discussions.

A support of the Slovak Grant Agency for Sciences VEGA, grant No.2/0105/21, is acknowledged.


\begin{thebibliography}{99}

\leftskip=-5pt \vspace{-0.3truecm}

\bibitem{Pedlar} T.E.Pedlar et al, Phys. Rev. Lett. {\bf 95}, 261803 (2005).

\bibitem{Ablikim1} M.Ablikim et al, Phys. Lett. {\bf B630}, 14 (2005).

\bibitem{Ablikim2} M.Ablikim et al, Phys. Rev. {\bf D91}, 112004 (2015).

\bibitem{Akhmetshin1} R.R.Akhmetshin et al, Phys. Lett. {\bf B759}, 634 (2016).

\bibitem{Akhmetshin2} R.R.Akhmetshin et al, Phys. Lett. {\bf B794}, 64 (2019).

\bibitem{Ablikim3} M.Ablikim et al, Phys. Rev. Lett. {\bf 124},  042001 (2020)

\bibitem{Aubert} B.Aubert et al, Phys. Rev {\bf D73}, 012005 (2006)b b

\bibitem{Lees1} J.P.Lees et al, Phys. Rev. {\bf D87}, 092005 (2013).

\bibitem{Lees2} J.P.Lees et al, Phys. Rev. {\bf D88}, 032011 (2013).

\bibitem{Ablikim4} M.Ablikim et al., Phys. Rev. {\bf D99}, 092002 (2019).

\bibitem{Ablikim5} M.Ablikim et al., Phys. Lett. {\bf B817}, 136328 (2021).

\bibitem{BaPaZa} R.Baldini Ferroli, S Paceti, A.Zalo, Eur. Phys. J. {\bf A48}, 33 (2012).

\bibitem{TomRek} E.Tomasi-Gustafsson, M.Rekalo, Phys. Lett. {\bf B504}, 291 (2001).

\bibitem{BianTom} A.Bianconi, E.Tomasi-Gustafsson, Phys. Rev. Lett. {\bf 114}, 232301 (2015).

\bibitem{T-GP} E.Tomasi-Gustafsson, S.Paceti, Phys. Rev. {\bf C106}, 035203 (2022).

\bibitem{BDD} E.Bartos, S.Dubnicka, A.Z.Dubnickova, Dynamics {\bf 3}, 137 (2023).

\bibitem{Lees3} J.P.Lees et al, Phys. Rev. {\bf D88}, 032013 (2013).

\bibitem{Lees4} J.P.Lees et al, Phys. Rev. {\bf D92}, 072018 (2015).

\bibitem{Ablikim6} M.Ablikim et al., Phys. Rev. {\bf D99}, 032001 (2019).

\bibitem{DD} S.Dubnicka, A.Z.Dubnickova, Acta Phys. Slovaca {\bf Vol.60}, 1 (2010).

\bibitem{PDG} P.A.Zyla et al.(Particle Data Group) Prog. Theor. Exp. Phys. 2020, 083c01 (2020).

\bibitem{BDLDK} E.Bartos, S.Dubnicka, A.Liptaj, A.Z.Dubnickova, R.Kaminski, Phys. Rev. D 96, 113004 (2017).

\bibitem{Lees5} J.P.Lees et al, Phys. Rev. {\bf D89}, 092002 (2014).

\bibitem{Kozyrev} E.A.Kozyrev et al, Phys. Lett. {\bf B760}, 314 (2016).

\bibitem{Ablikim7} M.Ablikim et al., Phys. Rev. {\bf D104}, 092014 (2021).

\bibitem{Ablikim8} M.Ablikim et al., Nat. Phys. {\bf 17}, 1200 (2021).



\end{thebibliography}
\end{document}